\documentstyle[12pt,epsfig]{article}
\advance\textheight by \topskip
\begin{document}
\tolerance=100000
\thispagestyle{empty}
\setcounter{page}{0}

\newcommand{\be}{\begin{equation}}
\newcommand{\ee}{\end{equation}}
\newcommand{\br}{\begin{eqnarray}}
\newcommand{\er}{\end{eqnarray}}
\newcommand{\ba}{\begin{array}}
\newcommand{\ea}{\end{array}}
\newcommand{\bi}{\begin{itemize}}
\newcommand{\ei}{\end{itemize}}
\newcommand{\bn}{\begin{enumerate}}
\newcommand{\en}{\end{enumerate}}
\newcommand{\bc}{\begin{center}}
\newcommand{\ec}{\end{center}}
\newcommand{\ul}{\underline}
\newcommand{\ol}{\overline}
\newcommand{\ar}{\rightarrow}
\newcommand{\sm}{${\cal {SM}}$}
\newcommand{\mssm}{${\cal {MSSM}}$}
\newcommand{\susy}{{{SUSY}}}
\def\epem{\ifmmode{e^+ e^-} \else{$e^+ e^-$} \fi}
\newcommand{\Dir}{\kern -6.4pt\Big{/}}
\newcommand{\Dirin}{\kern -10.4pt\Big{/}\kern 4.4pt}
\newcommand{\DDir}{\kern -7.6pt\Big{/}}
\newcommand{\DGir}{\kern -6.0pt\Big{/}}
\newcommand{\eett}{$e^+e^-\rightarrow t\bar t $}
\newcommand{\eeZphi}{$e^+e^-\rightarrow Z\phi $}
\newcommand{\eeZH}{$e^+e^-\rightarrow ZH $}
\newcommand{\eeAH}{$e^+e^-\rightarrow AH $}
\newcommand{\eehww}{$e^+e^-\rightarrow hW^+W^-$}
\newcommand{\bbww}{$b\bar b W^+W^-$}
\newcommand{\eebbww}{$e^+e^-\rightarrow b\bar b W^+W^-$}
\newcommand{\ttbbww}{$t\bar t\rightarrow b\bar b W^+W^-$}
\newcommand{\eettbbww}{$e^+e^-\rightarrow t\bar t\rightarrow b\bar b W^+W^-$}
\newcommand{\bbb}{$ b\bar b$}
\newcommand{\ttb}{$ t\bar t$}
\def\Ord{\buildrel{\scriptscriptstyle <}\over{\scriptscriptstyle\sim}}
%
\def\OOrd{\buildrel{\scriptscriptstyle >}\over{\scriptscriptstyle\sim}}
\def\Ord{\buildrel{\scriptscriptstyle <}\over{\scriptscriptstyle\sim}}
\def\OOrd{\buildrel{\scriptscriptstyle >}\over{\scriptscriptstyle\sim}}
\def\pl #1 #2 #3 {{\it Phys.~Lett.} {\bf#1} (#2) #3}
\def\np #1 #2 #3 {{\it Nucl.~Phys.} {\bf#1} (#2) #3}
\def\zp #1 #2 #3 {{\it Z.~Phys.} {\bf#1} (#2) #3}
\def\pr #1 #2 #3 {{\it Phys.~Rev.} {\bf#1} (#2) #3}
\def\prep #1 #2 #3 {{\it Phys.~Rep.} {\bf#1} (#2) #3}
\def\prl #1 #2 #3 {{\it Phys.~Rev.~Lett.} {\bf#1} (#2) #3}
\def\mpl #1 #2 #3 {{\it Mod.~Phys.~Lett.} {\bf#1} (#2) #3}
\def\rmp #1 #2 #3 {{\it Rev. Mod. Phys.} {\bf#1} (#2) #3}
\def\sjnp #1 #2 #3 {{\it Sov. J. Nucl. Phys.} {\bf#1} (#2) #3}
\def\cpc #1 #2 #3 {{\it Comp. Phys. Comm.} {\bf#1} (#2) #3}
\def\xx #1 #2 #3 {{\bf#1}, (#2) #3}
\def\preprint{{\it preprint}}

\begin{flushright}
{RAL-TR-1999-001}\\
{January 1999}\\
\end{flushright}

\vspace*{\fill}

\begin{center}
{\Large \bf
Irreducible QCD backgrounds to top searches \\[0.5 cm]
in semi-leptonic final states\\[0.75 cm]
at the Next Linear Collider}\\[1.0 cm]
{\large S. Moretti}\\[0.75 cm]
{\it Rutherford Appleton Laboratory,}\\
{\it Chilton, Didcot, Oxon OX11 0QX, UK.}\\[0.5cm]
\end{center}
\vspace*{\fill}

\begin{abstract}
{\noindent 
\small
At future electron-positron colliders, one of the
largest irreducible backgrounds to top searches in the channel
`4 jets + lepton + missing energy' comes from QCD events of
order $\alpha_s^2$. 
We compute here such processes exactly at the parton level
by resorting to 2 $\ar$ 6 matrix elements exploiting helicity amplitude
techniques. We adopt a typical selection procedure
based on the tagging of a high momentum and separated lepton.
We finally outline kinematic differences between signal and background
events that can be exploited to further reduce such a QCD noise.}
\end{abstract}
\vskip4.0cm
\hrule
\vskip0.25cm
\noindent
Electronic mail: moretti@v2.rl.ac.uk.

\vspace*{\fill}
\newpage

\section{Introduction}
\label{sec:intro}

One of the {\sl top} on the list 
reasons to build an $e^+e^-$ linear collider (NLC) with a
centre-of-mass (CM) energy between 350 and 1 TeV \cite{ee500} is to study in
great detail the {\sl top} parameters: its mass ($m_t$), width ($\Gamma_t$),
quantum numbers ($Q_t$, $I^3_t$)  
and branching ratios (BRs) \cite{top}. Not surprisingly so, as 
it is not unreasonable to believe that the heaviest of the fundamental 
particles discovered so far would after all have 
something to teach us \cite{topthe}. Indeed, 
in the unforeseen scenario 
that no Higgs bosons and no Supersymmetric (SUSY) particles
are found at the LHC \cite{LHC}, this might even be the 
only task left for the NLC:
to run as a  top factory\footnote{The  other,
more frightening legacy of a Higgs-
and SUSY-less LHC would be a strongly interactive weak sector, building
up somewhere around the TeV scale, the very upper end of, if not beyond,
the technical reach of the NLC as well as of many 
perturbative calculations !}.
Needless to say, under such a gloomy prospect, one would want to make 
the most out of such a machine, one way or another. 

Our contribution in that respect is to calculate
the $2\ar6$ scattering processes
\begin{equation}\label{qqQQ}
e^+e^-\ar q\bar q~Q\bar Q'~\ell \nu_\ell
\end{equation} 
\begin{equation}\label{qqgg}
e^+e^-\ar g     g~Q\bar Q'~\ell \nu_\ell
\end{equation}
at the parton level, through the perturbative order
${\cal O}(\alpha_{em}^4\alpha_s^2)$, for any possible flavour combination
of quarks $q,Q,Q'=u, ... b\ne t$ 
and leptons $\ell,\nu_\ell$, with $\ell=e,\mu\ne\tau$. 
The sum of these two mechanisms represents one of  
the largest `irreducible' backgrounds to top production and decay in the
semileptonic channel
\begin{equation}\label{tt}
e^+e^-\ar t\bar t\ar b\bar b~Q\bar Q'~\ell \nu_\ell,
\end{equation} 
the one preferred for experimental studies \cite{topexp}.
Their calculation has never been attempted before\footnote{A preliminary
exercise in such direction was performed in \cite{io}, where
however only the case of on-shell $W^\pm$ production plus
four jets was considered. For the case of ${\cal O}(\alpha_{em}^6)$
backgrounds see Ref.~\cite{sandro}.}. Obviously,
if one wants to perform precision measurements of top parameters
at the NLC, to pin down the size and shape of all important background
processes is of paramount importance. 

The plan of the paper is as follows. Next Section describes our calculations.
The third one presents our results. Conclusions are in 
Sect.~\ref{sec:conc}.

\section{Calculation}
\label{sec:calc}

\begin{figure}[p]
~\epsfig{file=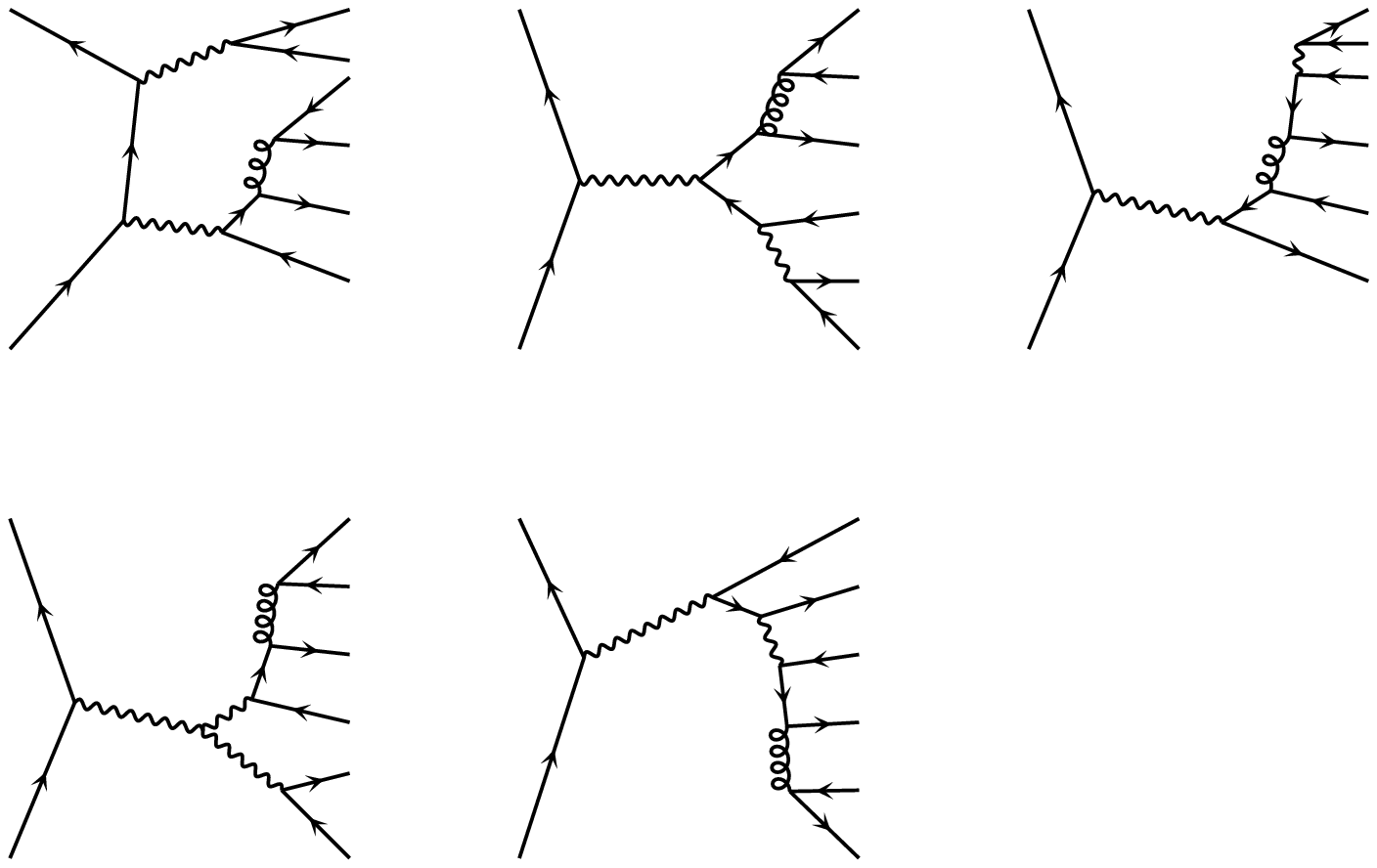,height=20cm}  
\vskip-6cm
\caption{Relevant Feynman sub-diagrams contributing at lowest
order to process (\ref{qqQQ}). Permutations of real and virtual
lines along the fermion lines are not shown. An internal wavy line
represents a $W^\pm$, a $\gamma$ or a $Z$, as appropriate.}
\label{fig:diagrams_qqQQ}
\end{figure}

\begin{figure}[p]
~\epsfig{file=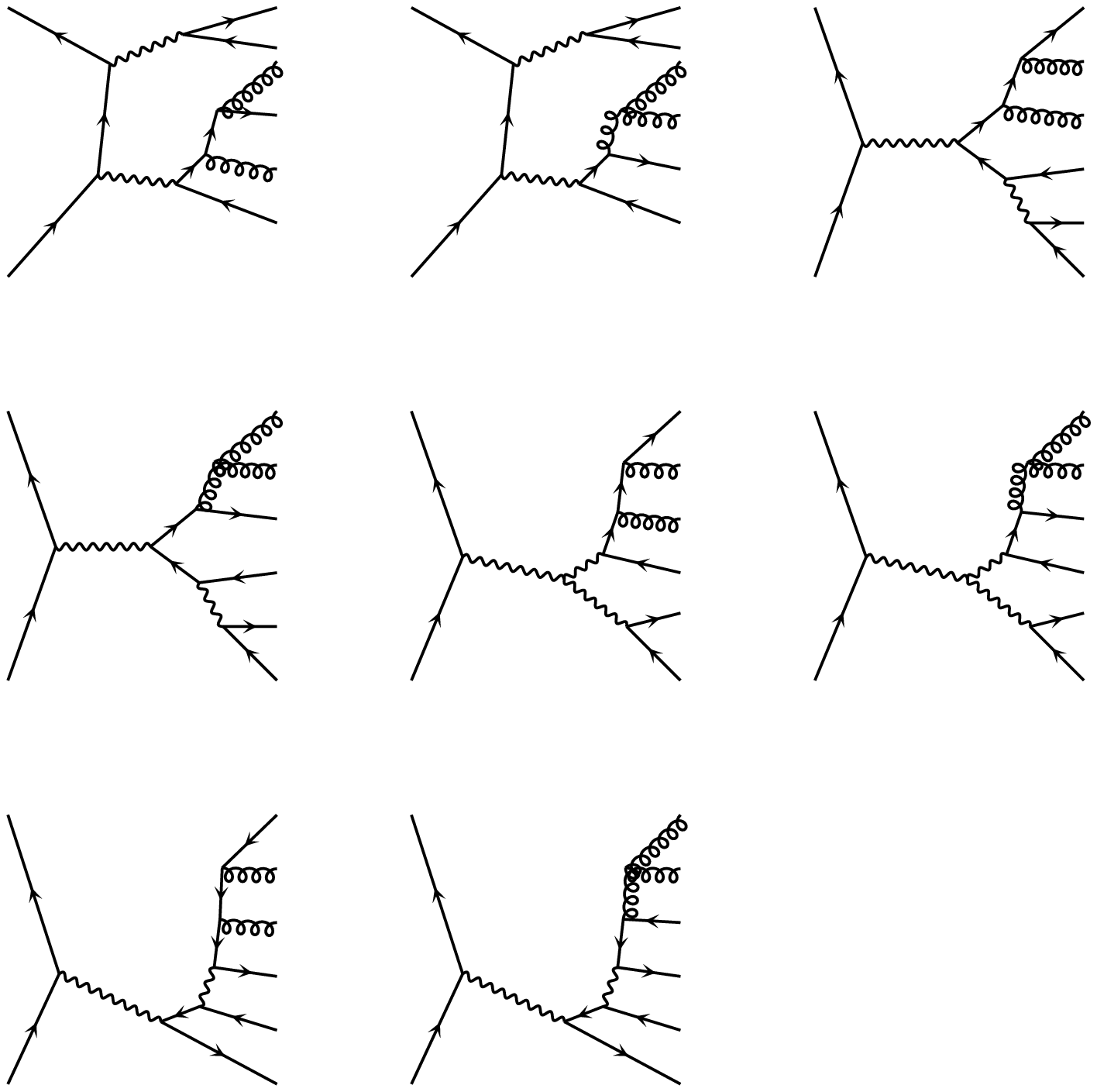,height=20cm}  
\vskip-6cm
\caption{Relevant Feynman sub-diagrams contributing at lowest
order to process (\ref{qqgg}). Permutations of real and virtual
lines along the fermion lines are not shown. An internal wavy line
represents a $W^\pm$, a $\gamma$ or a $Z$, as appropriate.}
\label{fig:diagrams_qqgg}
\end{figure}

Most of the Feynman diagrams that one 
has to tackle in order to calculate processes
(\ref{qqQQ})--(\ref{qqgg}) proceed through
$W^{\pm *}$ + 4 jet production \cite{io}, 
with the gauge bosons subsequently decaying
to lepton-neutrino pairs, see first four(six) graphs in 
Fig.~\ref{fig:diagrams_qqQQ}(\ref{fig:diagrams_qqgg}).
In addition to these, one also has to consider 
the graphs in which the off-shell $W^{\pm *}$
boson is produced in the bremsstrahlung off a leptonic current, 
see the last one(two) in 
Fig.~\ref{fig:diagrams_qqQQ}(\ref{fig:diagrams_qqgg}), and eventually
decays to four partons. 
Altogether one has to compute 140 diagrams: 32 associated with process
(\ref{qqQQ}) and 108 with (\ref{qqgg}). 
(Notice that for the time being we neglect diagrams in which an
electron and the companion neutrino in the final states of 
(\ref{qqQQ})--(\ref{qqgg}), i.e., $\ell=e$, are connected to the
incoming beams via so-called `multi-peripheral' channels. 
We will come back to this point in the conclusive Section.)
The signal comes via two simple
$s$-channel graphs (that we do not reproduce here).

We calculate the signal (\ref{tt}) at the leading-order (LO), though we
are aware that several higher order electroweak and QCD corrections (mainly 
to the on-shell production) are known to date \cite{topthe,bern}.
We do this for consistency, as the background processes 
(\ref{qqQQ})--(\ref{qqgg}) can only be evaluated at tree-level with
present technology. To compute all 
graphs  is not a prohibitive task, if one resorts to helicity
amplitudes methods. We have done so, by using both MadGraph \cite{tim}
and a self-made program based on the technique of Ref.~\cite{Berends}.
They agree with each other. Moreover, they have passed all
our gauge-invariance tests, so to give us confidence in our numerical 
results.

To obtain the latter, one has to integrate the Feynman amplitudes
squared over a six-body phase space. This task is not difficult either,
provided one takes some special care in dealing
 with the various resonances.
To get around this problem we have proceeded as described in 
Ref.~\cite{paps}. That is, by splitting the gauge-invariant
matrix elements of (\ref{qqQQ})--(\ref{qqgg})
in several sub-terms, each having its peculiar resonant 
structure. To any of these a dedicated mapping of the phase space 
has been attributed. These pieces have eventually been
 summed up together,
after integration (performed with different packages, 
for cross-check purposes), so to recover gauge-invariance. A flat phase
space has been instead used to integrate the
interferences between the different
sub-terms, with the help of some brute force too (i.e., a largely increased
number of random calls). In general, we have verified that 
their contribution is never
dominant, but not necessarily negligible, as compared to the
pure resonances. For reason of space, 
we will not dwell here in technicalities any further, as we will discuss
only the total integrated rates, summed over all production sub-channels.

Before proceeding to present our results, we list the numerical values
adopted for the various Standard Model parameters:
$$m_\ell=m_{\nu_\ell}=m_u=m_d=m_s=m_c=0,
$$
$$\qquad m_b=4.25~{\mathrm{GeV}},
\qquad m_t=175~{\mathrm{GeV}},$$
$$M_Z=91.175~{\mathrm {GeV}},\quad\quad \Gamma_Z=2.5~{\mathrm {GeV}},$$
$$M_W=80.23~{\mathrm {GeV}},\quad\quad \Gamma_W=2.08~{\mathrm {GeV}}.$$
As for the top width $\Gamma_t$, 
we have used the leading-order (LO) value of 1.55 GeV
as a default. However, in a few cases, we have compared the yield of process
(\ref{tt}) with finite width effects to that of the same 
reaction in Narrow Width
Approximation (NWA), for which we have rewritten the 
(denominator of the) top quark propagator
as    
\begin{equation}\label{NWA}
\frac{1}{p^2-m_t^2+{\mathrm{i}}m_t\Gamma}
\left(\frac{\Gamma}{\Gamma_t}\right)^{1/2},
\end{equation}
with $\Gamma\ar0$, 
thus mimicking a delta function\footnote{For 
$\Gamma\Ord 10^{-5}$ GeV 
the total cross sections in NWA are stable
and reproduce the on-shell 
results of the $2\ar2$ process $e^+e^-\ar t\bar t$ within
numerical accuracy. (Note that
for $\Gamma\equiv\Gamma_t$ in eq.~(\ref{NWA}) 
the standard expression of the propagator is recovered.)} $\delta(p^2-m_t^2)$.
  
For the vector and axial couplings
of the gauge bosons to the fermions, we use the `effective leptonic' value
\be\label{s2w_eff}
\sin^2(\theta_W)\equiv
\sin^2_{\mathrm {eff}}(\theta_W)=0.2320.
\ee
The strong coupling constant $\alpha_s$ entering
processes (\ref{qqQQ})--(\ref{qqgg}) has been evaluated
at two loops, with $N_f=5$ and $\Lambda_{\overline{\mathrm {MS}}}=200$ MeV,
at the scale $Q^2=s$. The electromagnetic one was set at $1/128$.
Finally, the centre-of-mass (CM) energies considered for the NLC are
$\sqrt s\equiv E_{\mathrm{cm}}=360$ and 500 GeV, 
as representative of the threshold,
$\sqrt s\OOrd2m_t$, and 
asymptotic, $\sqrt s\gg2m_t$, top-antitop production regimes.
Beyond those energies is no longer the top quark realm. 

\section{Results}
\label{sec:res}

In discussing the interplay between the background (\ref{qqQQ})--(\ref{qqgg})
and signal (\ref{tt}) processes, we have focused our attention to the case of
the semi-leptonic (or, equivalently, semi-hadronic) signature
\begin{equation}\label{sign}
4~{\mathrm{jets}} + \ell^\pm + E\DDir,
\end{equation}
where $\ell=e$ or $\mu$, $E\DDir$ represents the missing energy/momentum due
to the neutrino $\nu_\ell$ 
escaping detection and where to the four-jet hadronic system
no $b$-tagging is 
applied\footnote{If the latter is enforced, we have already shown in
Ref.~\cite{io} that the 
irreducible background to top events due to `$W^\pm + 2~b$ +
2 jet' events is negligible, 
provided that a vertex tagging efficiency $\epsilon_b\OOrd0.5$
can be achieved.}. 
We identify the quarks and gluons in (\ref{qqQQ})--(\ref{tt}) with 
the jets in (\ref{sign}) and apply all our cuts directly at the partonic level.

The choice of considering here only the final state (\ref{sign})
is indeed not restrictive, in the sense that the latter is to date the 
experimentally preferred channel in searching
for $e^+e^-\ar t\bar t\ar b\bar b W^+W^-$ events. On the one hand, as 
opposed to
the fully hadronic 
signature $b\bar b W^+W^-\ar 6$ jets, it has a simpler detector
topology and thus is much easier to reconstruct, further allowing one to reduce
the severe problems due to the six-jet combinatorics (particularly, if no heavy
quark identification is exploited)\footnote{Also notice that, once the missing 
energy/momentum 
has been assigned to the neutrino, the kinematics of (\ref{sign})
is fully constrained, like in the case of the purely hadronic channel.}. 
On the other hand, the case involving two leptonic $W^\pm$ decays has a double
disadvantage as compared to channel (\ref{sign}), that is, a very much reduced
statistics and problems in 
reconstructing the top mass spectra because of the two 
neutrinos. Furthermore,
as selection method of candidate top-antitop events we adopt one rather similar
to that outlined in Sect.~4.2 of  
Ref.~\cite{topexp}, based on the detection of 
high-momentum isolated leptons. As a matter 
of fact, such a procedure has been shown to be the most effective one, 
as it eventually yields the largest signal-to-background ratio, both at and
above threshold: see Tab.~4.2 of \cite{topexp}. 

\begin{figure}[p]
~\epsfig{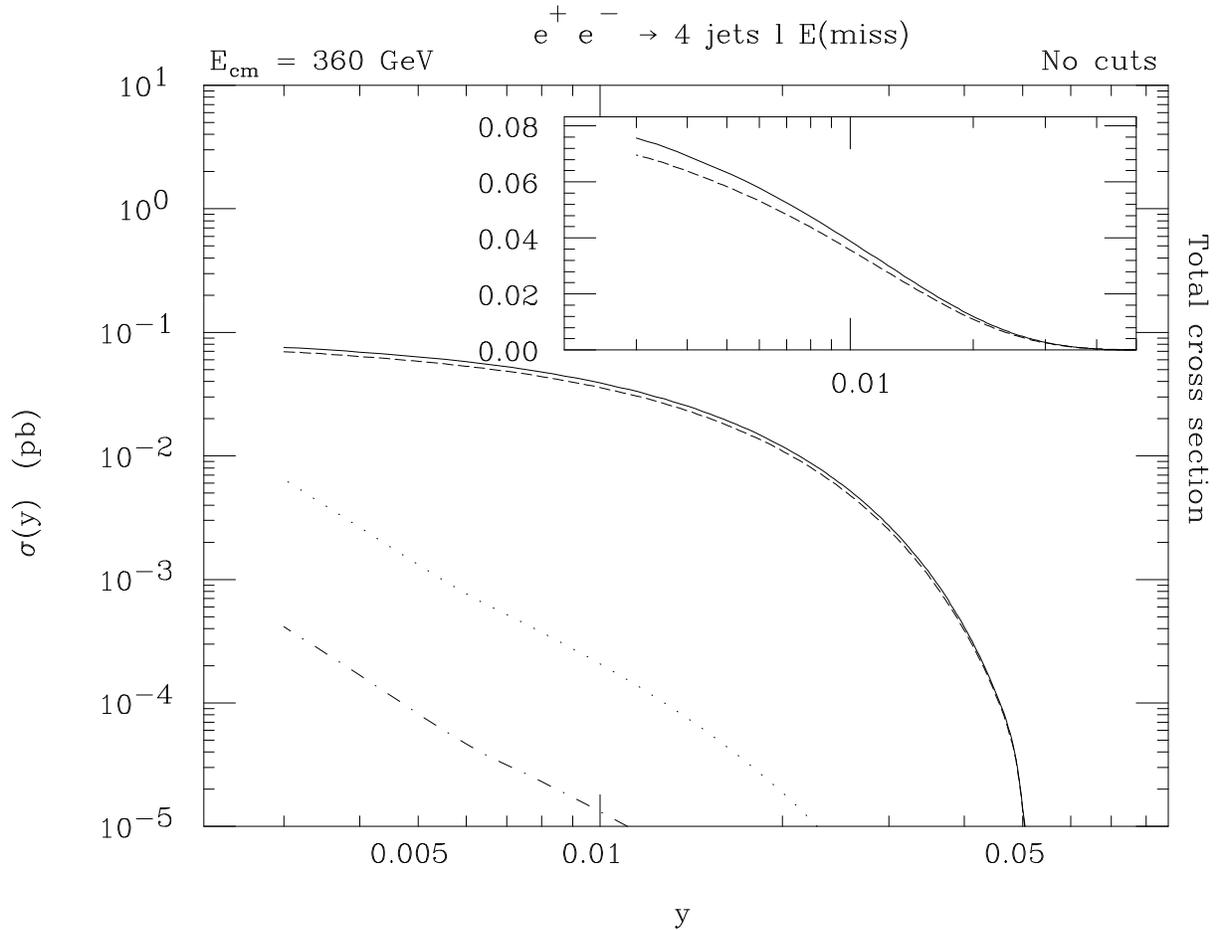}  
\vskip1cm
\caption{Cross sections of the processes:
(\ref{tt}) in NWA (solid) and with finite top width (dashed),
(\ref{qqgg}) (dotted) and (\ref{qqQQ}) (dot-dashed), as a function
of $y$ for the Jade algorithm (applied to the four jets
and the lepton), at $\protect{\sqrt s=360}$ GeV,
before the selection cuts (\ref{selection}) and without ISR.  
The insert refers to the top-antitop rates only (labelled as above).}
\label{fig:y360}
\end{figure}

\begin{figure}[p]
~\epsfig{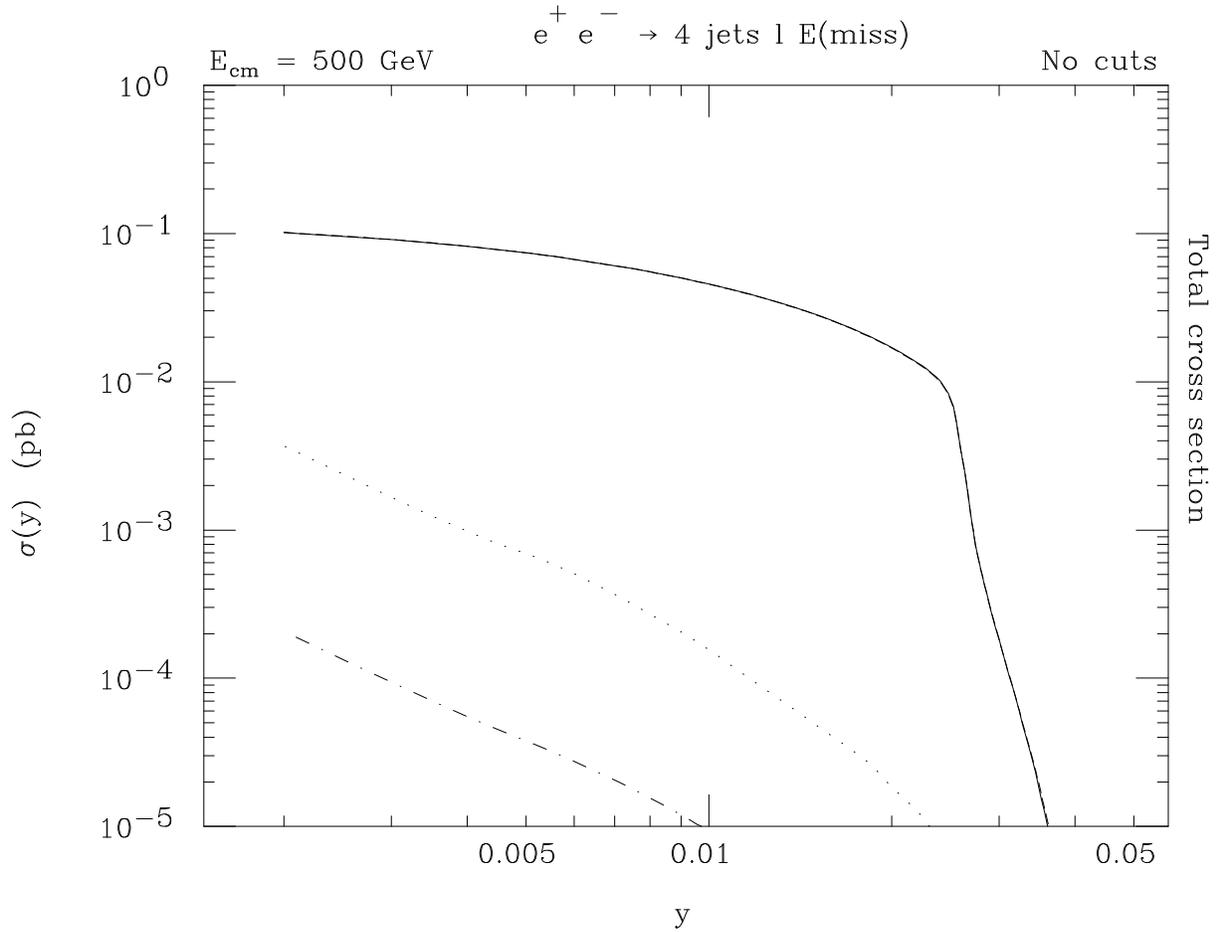}  
\vskip1cm
\caption{Cross sections of the processes:
(\ref{tt}) in both NWA and with finite top width (solid,
the two visually coincide),
(\ref{qqgg}) (dotted) and (\ref{qqQQ}) (dot-dashed), as a function
of $y$ for the Jade algorithm (applied to the four jets
and the lepton), at $\protect{\sqrt s=500}$ GeV,
before the selection cuts (\ref{selection}) and without ISR.}
\label{fig:y500}
\end{figure}

About $40\%$ of top-antitop events (\ref{tt}) 
produce an energetic electron or muon which 
is clearly separated from the hadronic system. 
Following Ref.~\cite{topexp}, we consider
a lepton to be isolated if a jet clustering algorithm with a `minimum 
mass' cut-off
recognises it as a `jet' with a single particle. 
As jet finder we use the Jade one 
\cite{JADE},
with $y>0.003(0.002)$ at $\sqrt s=360(500)$ GeV, so that 
$M_{\ell^\pm j}\equiv 2E_{\ell^\pm}E_j(1-\cos\theta_{\ell^\pm j})>19(22)$ GeV
for each jet $j$ in (\ref{sign}). As the
MEs of the background processes are divergent if the partons are allowed
to be infinitely soft and/or 
collinear, we also apply the jet clustering algorithm
to the hadronic part of the event, for all cases (\ref{qqQQ})--(\ref{tt}).
Fig.~\ref{fig:y360} 
presents the $y$-dependent total rates for these three processes as 
obtained by 
enforcing the jet clustering algorithm only, e.g., at $\sqrt s=360$ GeV.
Notice that there exists a hierarchy in the production rates:
\begin{equation}\label{hierarchy}
\sigma(e^+e^-\ar t\bar t\ar b\bar b~Q\bar Q'~\ell \nu_\ell)\gg
\sigma(e^+e^-\ar g     g~Q\bar Q'~\ell \nu_\ell)\gg
\sigma(e^+e^-\ar q\bar q~Q\bar Q'~\ell \nu_\ell).
\end{equation} 
At the minimum value of $y$ considered here, 
they approximately scale as 100:10:1.
If $\sqrt s=500$ GeV, see Fig.~\ref{fig:y500},
the relative ratio of process (\ref{tt}) to either of 
(\ref{qqgg}) or (\ref{qqQQ}) further increases, while that between the latter 
two suffers little from the CM energy scaling. As for top width effects, 
whereas
these are naturally sizable at threshold (with differences of about $10\%$,
see the top-right insert in Fig.~\ref{fig:y360}), 
they instead fall to the percent level
in the asymptotic regime (Fig.~\ref{fig:y500}) \cite{noitt}.  

In Figs.~\ref{fig:y360}--\ref{fig:y500} 
we have neglected considering Initial State 
Radiation (ISR) \cite{ISR}, that is, the 
presence of bremsstrahlung photons emitted 
by the incoming electron/positron 
beams\footnote{Also beamsstrahlung and Linac energy
spread \cite{ISR} in principle affect processes (\ref{qqQQ})--(\ref{tt}). 
In practise, for narrow beam designs of the NLC, their effects are much 
smaller 
as compared to those induced by ISR \cite{ISR}, 
so for the time being we neglect the former in our analysis.}. 
The main consequence of ISR is to lower the effective
CM energy, thus ultimately reducing(enhancing) the production rates of 
processes
whose cross sections increase(decrease) with $\sqrt s$. One thus expects
the top-antitop rates (\ref{tt}) 
to be rather sensitive to ISR, for two reasons. Firstly,
because of the $s$-channel 
topology of the Feynman diagrams involved (which tends
to increase the rates). Secondly,
at threshold, because the difference $\sqrt s-2m_t$ starts 
approaching the edge of the phase space (thus decreasing the rates). 
In contrast, the background rates 
(\ref{qqQQ})--(\ref{qqgg}) should depend much less on the ISR. On the one hand,
they are not purely $s$-channel. On the other hand, $\sqrt s$ is well 
above the 
heavy particle thresholds which can onset there (such as the dominant
$W^+W^-$). 

This dynamics
can be appreciated by comparing the total cross sections in the upper lines
of Tab.~\ref{tab:sigma}, when no selection cuts 
are applied apart from the jet clustering
algorithm. There, notice that 
the background rates (third and fourth column)
outside and inside brackets are rather steady,
at both collider energies. 
For signal events, both in NWA and with finite top width
(first and second column),
differences are much more sizable. 
At threshold, it is clearly the phase space suppression
to dominate, depleting the signal rates by up to $43\%$ (at the minimum $y$).
In the asymptotic regime, where phase space effects become negligible, 
the $s$-channel increase is overturned 
by the presence of the invariant mass constraints, as
also typical energies of the final state particles 
diminish because of ISR. 

Indeed, in presence of the latter, the response of processes 
(\ref{qqQQ})--(\ref{tt}) to the 
implementation of any selection cuts is no longer straightforward, as a 
consequence 
of the fact that ISR also induces a smearing of the differential 
distributions.
Thus, from now onwards, all 
our results will include initial state bremsstrahlung. 
(We will keep those without it only for reference purposes.) 
Among the various ways of implementing the ISR, we have adopted here the 
so-called Electron 
Structure Function (ESF) approach, based on the formulae given in
Ref.~\cite{Nicro}. In addition, hereafter, we will stop considering
the case of process (\ref{tt}) in NWA.

We now proceed by applying all other selection cuts of \cite{topexp} 
that can be
exploited at parton level too. Namely, an event is accepted if: 
\begin{enumerate}
\item its `thrust' (calculated
by using the four jet and the lepton momenta), $T_{\ell^\pm,~j}$, is  
significantly far from the infrared region typical of QCD events; 
\item the invariant mass of the hadronic 
system, $M_{4j}$, is far above the typical
resonances of background events ($M_W$ in our case);
\item the amount 
of missing energy, $E\DDir$, is rather contained, as in top events this is
typically less than ${\sqrt{m_t^2-M_W^2}}$;
\item  the (absolute) momentum of the isolated lepton, 
$|\vec{p}_{\ell^\pm}|$, is
above a minimum energy threshold 
and below a maximum one of standard acceptance.   
\end{enumerate}
Numerically, to account 
for other sources of background too, other than (\ref{qqQQ})--(\ref{qqgg}),
we require \cite{topexp}:
\begin{eqnarray}\label{selection}
T_{\ell^\pm,~j}<0.75
& \; &
M_{4j}>0.4{\sqrt s} \nonumber \\
E\DDir<0.4{\sqrt s}
& \; &
0.04{\sqrt s}<|\vec{p}_{\ell^\pm}|<0.3{\sqrt s}.
\end{eqnarray}
The second line in Tab.~\ref{tab:sigma} reports our findings.
After the cuts in (\ref{selection}) are enforced, 
the background from processes (\ref{qqQQ})--(\ref{qqgg})
amounts to about $1.6\%$ of the signal (\ref{tt}) at threshold, 
whereas well above that
the corresponding figure is $\approx0.9\%$. 
Thus, in both cases, the QCD noise is
under control.

\begin{figure}[p]
~\epsfig{file=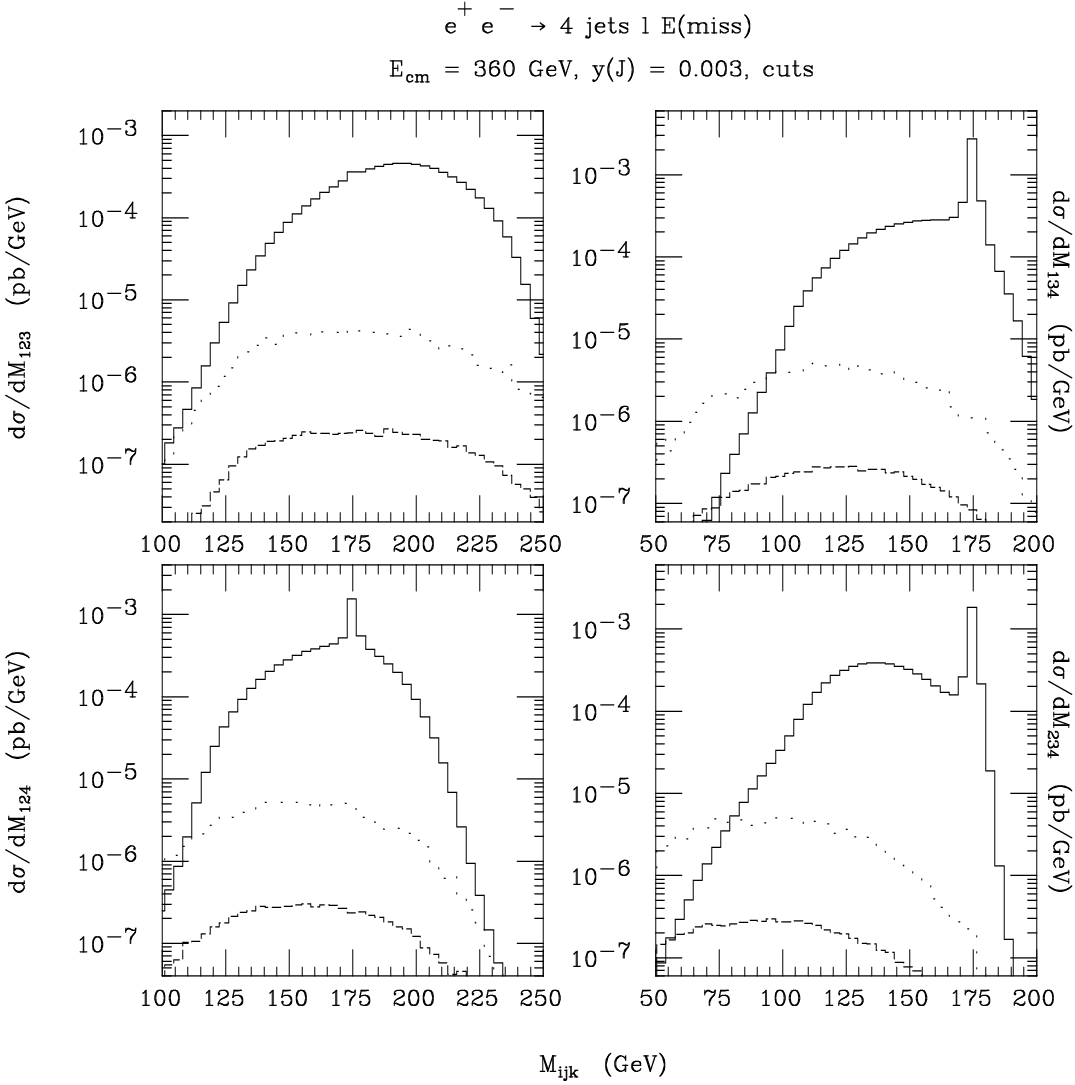,height=16cm,angle=0}  
\vskip1cm
\caption{Differential distributions in the invariant mass of
the energy-ordered three-jet pairs, $M_{ijk}$ with $i<j<k=1, ... 4$,
for the processes:
(\ref{tt}) with finite top width (solid),
(\ref{qqgg}) (dotted) and (\ref{qqQQ}) (dashed), 
for $y=0.003$ in the Jade algorithm (applied to the four jets
and the lepton), at $\protect{\sqrt s=360}$ GeV.
The selection cuts (\ref{selection}) have been enforced and
the ISR implemented.}
\label{fig:mjjj360}
\end{figure}

\begin{figure}[p]
~\epsfig{file=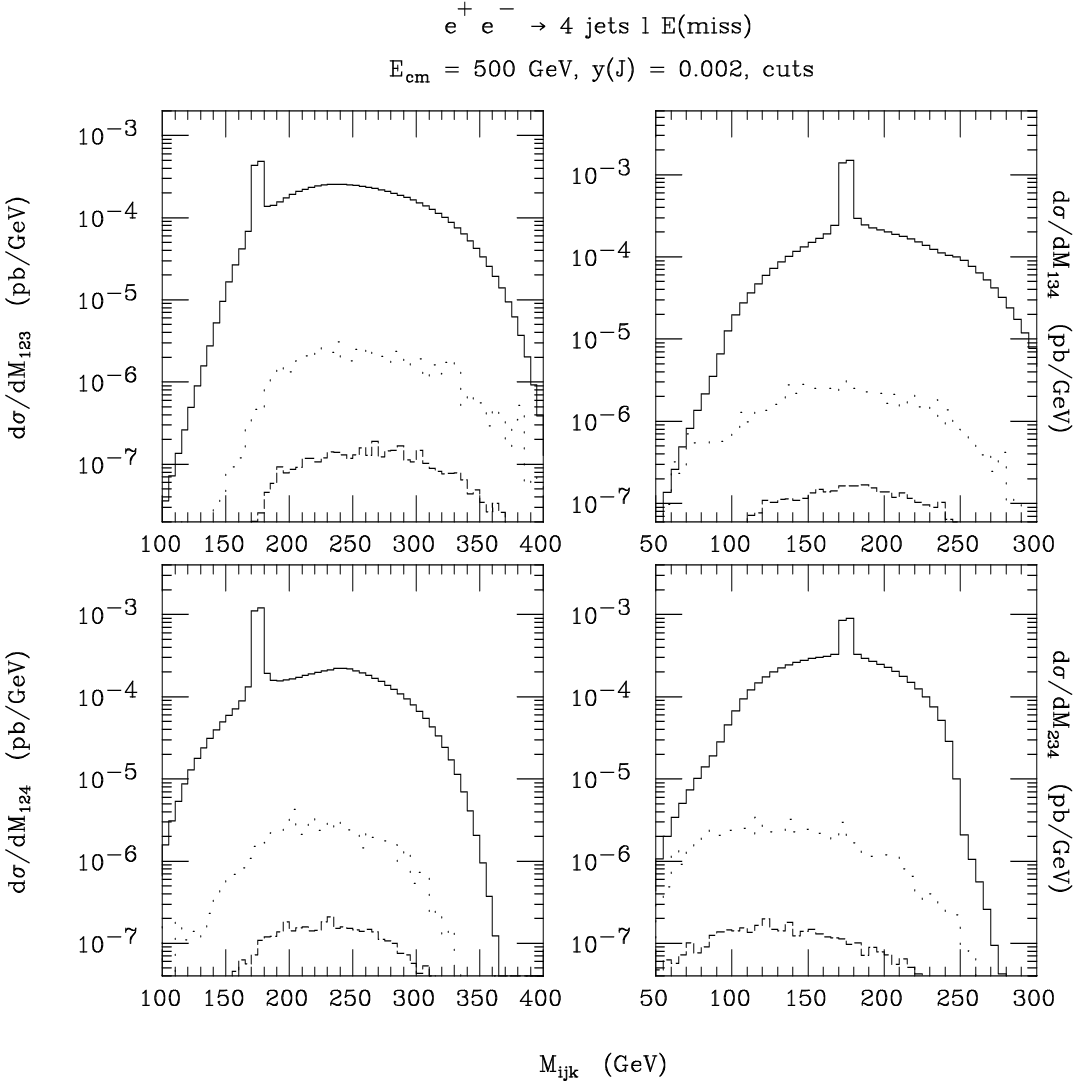,height=16cm,angle=0}  
\vskip1cm
\caption{Differential distributions in the invariant mass of
the energy-ordered three-jet pairs, $M_{ijk}$ with $i<j<k=1, ... 4$,
for the processes:
(\ref{tt}) with finite top width (solid),
(\ref{qqgg}) (dotted) and (\ref{qqQQ}) (dashed), 
for $y=0.002$ in the Jade algorithm (applied to the four jets
and the lepton), at $\protect{\sqrt s=500}$ GeV.
The selection cuts (\ref{selection}) have been enforced and
the ISR implemented.}
\label{fig:mjjj500}
\end{figure}

Nonetheless, one ought to know its effects on the differential spectra 
used to
fit the top parameters. 
We consider here all possible three-jet mass distributions 
$M_{ijk}$ which can 
be reconstructed in samples of the type (\ref{sign}). After ordering
the four jets in energy, such that $E_1 > ... > E_4$, one can build up 
four $ijk$ combinations, such that $i<j<k$. They are presented in
Figs.~\ref{fig:mjjj360} 
and \ref{fig:mjjj500}, for the cases $\sqrt s=360$ and 500
GeV, respectively. 
The size and
shape of the backgrounds (\ref{qqQQ})--(\ref{qqgg}) are rather
innocuous in the 
vicinity of the top peaks, so that one should not expect any significant
distortion of the Breit-Wigner distribution
of the top resonances. As a matter of fact, 
in this
respect, it is the intrinsic background due to mis-assigned jets originating 
in the signal (\ref{tt}) 
from $b$-quarks that affects most the signal, as discussed in 
Refs.~\cite{paps,noitt}. 

\begin{figure}[p]
~\epsfig{file=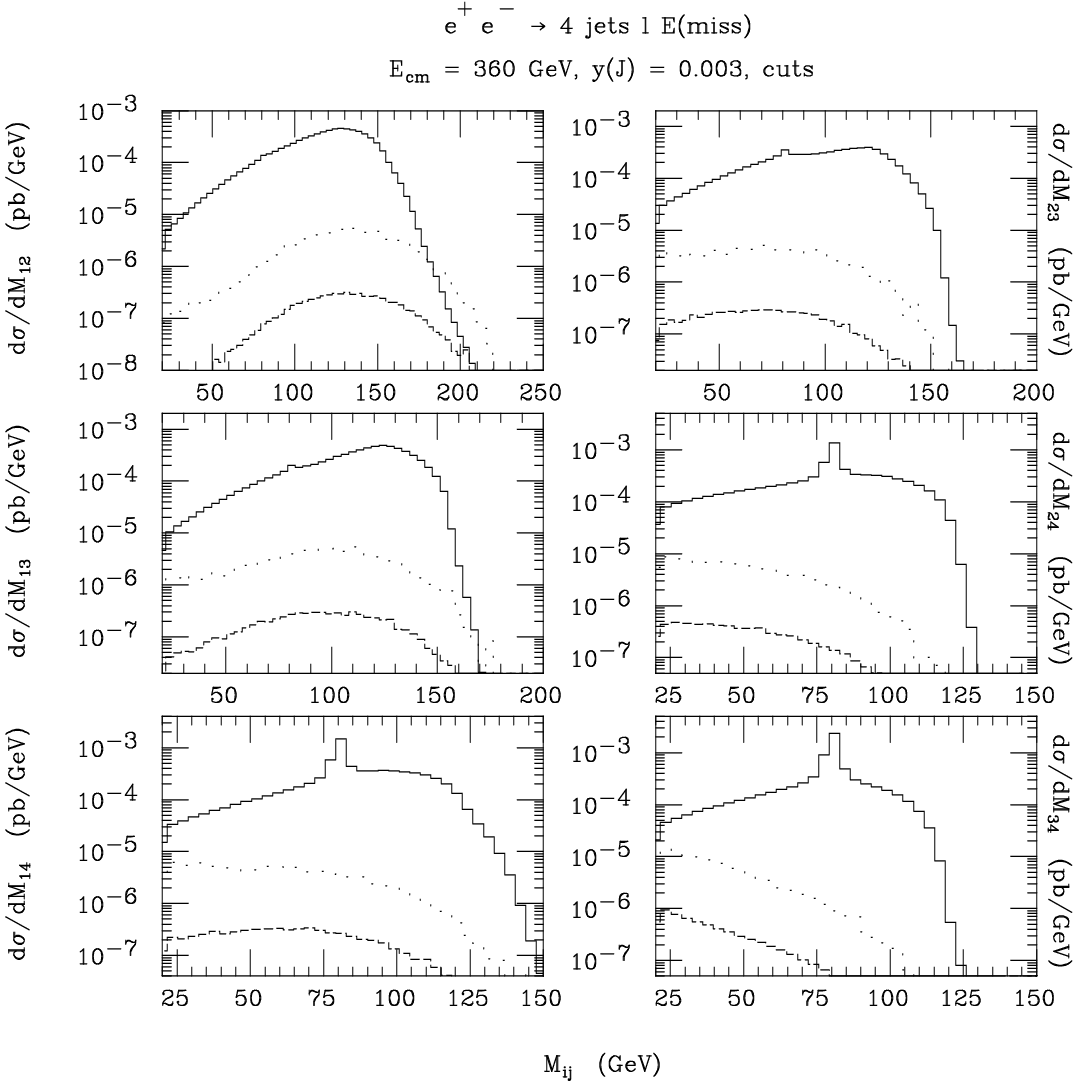,height=16cm,angle=0}  
\vskip1cm
\caption{Differential distributions in the invariant mass of
the energy-ordered two-jet pairs, $M_{ij}$ with $i<j=1, ... 4$,
for the processes:
(\ref{tt}) with finite top width (solid),
(\ref{qqgg}) (dotted) and (\ref{qqQQ}) (dashed), 
for $y=0.003$ in the Jade algorithm (applied to the four jets
and the lepton), at $\protect{\sqrt s=360}$ GeV.
The selection cuts (\ref{selection}) have been enforced and
the ISR implemented.}
\label{fig:mjj360}
\end{figure}

\begin{figure}[p]
~\epsfig{file=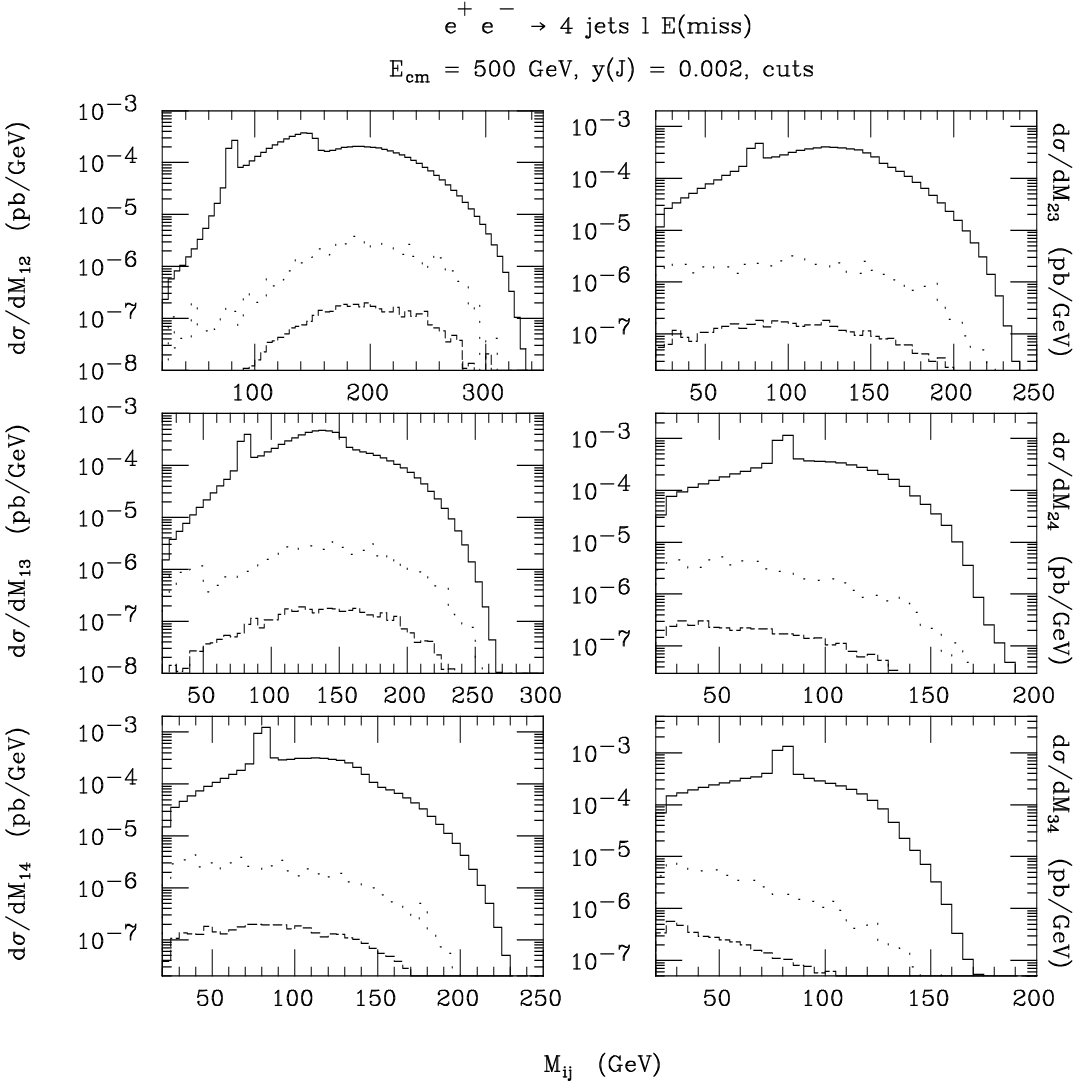,height=16cm,angle=0}  
\vskip1cm
\caption{Differential distributions in the invariant mass of
the energy-ordered two-jet pairs, $M_{ij}$ with $i<j=1, ... 4$,
for the processes:
(\ref{tt}) with finite top width (solid),
(\ref{qqgg}) (dotted) and (\ref{qqQQ}) (dashed), 
for $y=0.002$ in the Jade algorithm (applied to the four jets
and the lepton), at $\protect{\sqrt s=500}$ GeV.
The selection cuts (\ref{selection}) have been enforced and
the ISR implemented.}
\label{fig:mjj500}
\end{figure}

Finally notice that an additional requirement can be imposed to events of the
form (\ref{sign}), in order to increase the signal-to-background ratio
of (\ref{tt}) vs.~(\ref{qqQQ})--(\ref{qqgg}). 
That is,
that one two-jet combination $ij$, 
among the six possible possible ones, when $i<j=1,...4$, 
produces an invariant mass
$M_{ij}$ around the $W^\pm$ mass. As one can appreciate in 
Figs.~\ref{fig:mjj360}--\ref{fig:mjj500}, 
this would reduce the QCD noise to imperceptible
levels. If one imposes at $\sqrt s=360$ GeV, e.g., 
$|M_{34}-M_W|\le20$ GeV, than additional reduction factors 
of $9.1$ and $8.4$ apply to the processes (\ref{qqQQ}) and (\ref{qqgg}),
respectively, whereas the loss on the
signal (\ref{tt}) is just $1.3$. Corresponding numbers
at $\sqrt s=500$ GeV are $4.3$ and $3.7$ for the backgrounds,
and $1.6$ for the signal.
Finally, we should mention that
we have tried other quantities too (such as, e.g., jet energies,
relative angles, etc.) but they have proved themselves much less useful 
than the $M_{ij}$ spectra in disentangling
reactions (\ref{qqQQ})--(\ref{qqgg}) and (\ref{tt}).

\section{Conclusions}
\label{sec:conc}

Thus we conclude that irreducible ${\cal O}(\alpha_{em}^4\alpha_s^2)$ 
backgrounds to the 
`4 jets + lepton + missing energy' signature of top-antitop events
at the Next Linear 
Collider are reduced at the ten percent level by using a standard 
selection procedure (at 
typical design energies), in line with
previous results obtained by using parton
shower models. Such figure can vigorously be reduced 
further if a simple mass 
requirement on a two-jet system is imposed. We have obtained such 
results by computing tree-level  
matrix elements at leading-order for the relevant
$2\ar 6$ processes, with the only exception of multi-peripheral graphs
entering final states including electrons. 
We have neglected the letter for two reasons.
On the one hand, we would have had to calculate twice as many diagrams 
as compared to the way we did it.
On the other hand, the contributions of the missing terms
has already been proved to be
very small in the case of ${\cal O}(\alpha_{em}^6)$ 
$e^+e^-\ar b\bar b~Q\bar Q'~\ell \nu_\ell$ processes, as their inclusion
account for an increase of only $6\%$ with respect to the muon rates, 
for a CM energy of 500 GeV \cite{sandro}. Indeed, we expect the same
to have occurred here. Anyhow, given the results we have eventually
obtained  for the signal-to-background
ratio, even if their rate is actually much larger than the mentioned figure,
our conclusions would remain unchanged. Finally, we believe that,  
although confined at the partonic stage, our findings should not
be invalidated by 
studies at the hadronic level. We make our programs available to the
public for simulation purposes in the above respect.

\subsection*{Acknowledgements}

We thank the Department of Theoretical Physics of Granada University
for the kind hospitality while this project was conceived and
Ramon Mu\~noz-Tapia fur discussions.
Financial support is provided by the UK PPARC.

\goodbreak

\clearpage

\begin{table}[htb]
\begin{center}
\begin{tabular}{|c|c|c|c|}

\hline

\multicolumn{4}{|c|}
{$e^+e^-\ar 4~\mbox{jets}~+~\ell^\pm + E\DDir$ at the NLC}
\\ \hline\hline

\multicolumn{4}{|c|}
{$\sigma_{\mathrm{tot}}$ (fb)}
\\ \hline\hline

$t\bar t$ (NWA)             & 
$t\bar t\ar b\bar b W^+W^-$ & 
$W^\pm q\bar Qgg$           & 
$W^\pm q\bar Q q'\bar q'$   \\ 
\hline\hline

\multicolumn{4}{|c|}
{$\sqrt s=360$ GeV, 
$y^{\mathrm{J}}_{\ell^\pm,~j}>0.003~(M_{ij}>19~\mbox{GeV})$}
\\ \hline
$76(53)$ & $70(49)$ & $6.5(6.4)  $ & $0.42(0.42)  $ \\
$40(28)$ & $37(26)$ & $0.50(0.38)$ & $0.024(0.023)$ \\
\hline\hline

\multicolumn{4}{|c|}
{$\sqrt s=500$ GeV, 
$y^{\mathrm{J}}_{\ell^\pm,~j}>0.002~(M_{ij}>22~\mbox{GeV})$}
\\ \hline
$102(99)$ & $102(99)$ & $3.7(3.7)$ & $0.21(0.20)$   \\
$32(36)$ & $33(36)$ & $0.55(0.32)$ & $0.025(0.019)$ \\
\hline\hline

\multicolumn{4}{|c|}
{$T_{\ell^\pm,~j}<0.75$     \quad\quad\quad\quad\quad\quad\quad\quad
 $M_{4j}>0.4{\sqrt s}$} \\

\multicolumn{4}{|c|}
{$E\DDir<0.4{\sqrt s}$       \quad\quad\quad\quad
 $0.04{\sqrt s}<|\vec{p}_{\ell^\pm}|<0.3{\sqrt s}$}
\\ \hline\hline

\multicolumn{4}{|c|}
{No $b$-quark tagging}
\\ \hline

\end{tabular}
\end{center}
\vskip1.0cm
\caption{Cross sections of processes (\ref{qqQQ})--(\ref{tt}),
the latter in both NWA and with finite width, at $\sqrt s=360$ and
500 GeV.
First line is without the kinematical cuts (\ref{selection}). 
Second one is with the latter implemented.
In brackets, the same rates in presence of Initial State Radiation.
A default jet clustering procedure is enforced in all cases.}
\label{tab:sigma}
\end{table}

\clearpage

\vfill
\clearpage

\vfill
\clearpage

\vfill
\clearpage

\vfill
\clearpage

\vfill
\clearpage

\vfill
\clearpage

\vfill
\clearpage

\vfill

\end{document}